\documentstyle[aas2pp4,twoside,aabib]{article}
\input epsf
\def\s2n{S^{\prime}/N}
\def\vecr{{\bf r}}
\def\vecdr{{\bf \Delta r}}
\def\dr{{\Delta r}}

\def\vecr{{\bf r}}
\def\vecdr{{\bf \Delta r}}
\def\dr{{\Delta r}}

\def\inlinefig{1}

\begin{document}
\title{A New Method to Measure and Map the Gas Scale--Height of Disk Galaxies}

\author{Paolo Padoan\footnote{ppadoan@cfa.harvard.edu},
Sungeun Kim\footnote{skim@cfa.harvard.edu},
Alyssa Goodman\footnote{agoodman@cfa.harvard.edu}}
\affil{Harvard-Smithsonian Center for Astrophysics, Cambridge, MA 02138}
\author{and Lister Staveley--Smith}
\affil{Australia Telescope National Facility, CSIRO, PO Box 76, Epping, NSW 1710, Australia}

\begin{abstract}

We propose a new method to measure and map the gas scale height of nearby disk
galaxies. This method is applied successfully
to the Australia Telescope Compact Array interferometric HI survey of the Large 
Magellanic Cloud (LMC); it could also be applied to a significant number of nearby 
disk galaxies, thanks to the next generation of interferometric facilities, such as 
the extended VLA and CARMA.  

The method consists of computing the Spectral Correlation Function (SCF) for a  
spectral--line map of a face--on galaxy. The SCF quantifies the correlation
between spectra at different map positions as a function of their separation,
and is sensitive to the properties of both the gas mass distribution and the
gas velocity field. It is likely that spatial correlation properties of the gas
density and velocity fields in a galactic disk are sensitive to the value of the 
scale height of the gas disk. A scale--free turbulent cascade is unlikely to 
extend to scales much larger than the disk scale height, as the disk dynamics 
on those larger scales should be dominated by two dimensional motions. 

We find a clear feature in the SCF of the LMC HI disk, on the scale of 
$\approx 180$~pc, which we identify as the disk scale height. We are 
also tentatively able to map variations of the scale height over the disk.

\end{abstract}

%
%

\section{Introduction}

The gas scale height of galactic disks is an important quantity related to 
the process of star formation. It can be used to constrain the surface 
density of matter, the balance between star formation processes and gravity, 
and the ISM pressure, which must be important for regulating star formation.

Optical images of edge--on 
galaxies show that the scale height of stellar disks are independent of 
position along their major axis (Van der Kruit \& Searle 1981, 1982; Kylafis 
\& Bahcall 1987; Shaw and Gilmore 1990; Barnaby \& Thronson 1992), or 
slowly increasing with radius (de Grijs \& Peletier 1997). 
The constancy of the stellar scale height must be related to the radial and 
vertical distribution of the gas, since molecular clouds (MCs) are responsible
for both the formation of stars and the dynamical heating of the stellar disk
(Spitzer \& Schwarzschild 1951).

\nocite{vanderKruit+Searle81} \nocite{vanderKruit+Searle82} 
\nocite{Kylafis+Bahcall87} \nocite{Shaw+Gilmore90} \nocite{Barnaby+Thronson92} 
\nocite{deGrijs+Peletier97} \nocite{Spitzer+Schwarzschild51}

The vertical distribution of neutral hydrogen in our Galaxy has been studied
by many authors. It is found that the HI scale height $h$ is approximately 
constant over a large range of galactocentric radii (Dickey \& Lockman 1990; 
Heiles 1991). It is commonly assumed that the HI gas is supported in the 
Galactic gravitational potential primarily by turbulence (Lockman \& Gehman 1991), 
with contributions from magnetic and cosmic ray pressure (Parker 1966; Spitzer 1990) 
and radiation (Ferrara 1993). 

\nocite{Dickey+Lockman90} \nocite{Heiles91} \nocite{Lockman+Gehman91}
\nocite {Parker66} \nocite{Spitzer90} \nocite{Ferrara93}

The constancy of the gas scale height in galactic disks is a common assumption
in several models. However, it should not be taken for granted, since direct 
measurements of the HI scale height in external galaxies are difficult and 
sparse, and the HI scale height has never been mapped in a face--on galaxy. 
A nearby edge-on galaxy where the HI distribution has been studied with 
very high linear resolution is NGC 891. Sancisi \& Allen (1979) found the 
HI disk of NGC 891 to be less than 1~kpc thick near the center, increasing 
outwards up to 1--2 kpc. The HI disk flare is not confirmed in the later 
study by Swaters, Sancisi \& van der Hulst (1997), where a significant HI 
halo component is instead identified, and it is pointed out that the inferred 
HI vertical density profile is dependent on the adopted kinematic model.    
The CO scale height for the same galaxy ($\approx 220$~pc) has also been 
estimated (Scoville et al. 1993).

\nocite{Sancisi+Allen79} \nocite{Swaters+97} \nocite{Scoville+93}

In this Letter we propose a new method to measure and map the HI scale height 
of face--on galaxies. The method is based on the application of the Spectral 
Correlation Function (SCF) to HI interferometric surveys. The SCF 
(Rosolowsky et al. 1999; Padoan, Rosolowsky \& Goodman 2001) quantifies 
the correlation between spectra at different map positions as a function of 
their separation, and is sensitive to the properties of both 
the gas mass distribution and the gas velocity field. The scale 
corresponding to the scale height of the HI disk should mark the 
transition from small--scale three--dimensional turbulence to two--dimensional 
large--scale motions on the disk, since the ratio between radius and scale 
height of the cold gas in typical disk galaxies is very large ($\sim 100$). 
This transition should appear as a feature in the SCF at a spatial lag 
approximately equal to the HI scale height. On smaller scales the SCF 
should instead be a power law reflecting the self--similar nature of 
turbulence, as in MCs (Padoan, Rosolowsky \& Goodman 2001).

\nocite{Rosolowsky+99} \nocite{Padoan+2001SCF}

We have applied this method to the Australia Telescope Compact Array (ATCA) 
interferometric HI survey of the Large Magellanic Cloud (LMC) by Kim et 
al. (1998a,b, 1999), combined with the single dish survey done with the 
Parkes Multibeam by Staveley--Smith et al. (in preparation). Results of 
the combined survey are presented in Kim et al. (in preparation --see 
also astro--ph/0009299). We find that the SCF is indeed a power law 
on small scale, and has a characteristic feature at approximately 
180~pc, where the power law is interrupted. We identify the scale 
of the clear ``break'' in the SCF as the HI scale height. We are also 
able to map significant variations of the scale height over the disk, 
between $\approx 140$ and 240~pc.

\nocite{Kim+98a} \nocite{Kim+98b} \nocite{Kim+99}

In a recent work, Elmegreen, Kim \& Staveley--Smith (2001) have computed 
the power spectrum of HI intensity maps of the LMC from the same survey 
used in this work. They find that the power spectrum of intensity is 
steeper on small scales than on large ones. The transition occurs at 
approximately 80--100~pc, and is interpreted as the line of sight 
thickness of the cold HI layer. 
\nocite{Elmegreen+2001LMC} \nocite{Lazarian+Pogosian2000}
The present Letter is qualitatively a confirmation of this earlier result, 
in the sense that both works show the possibility of using the statistical 
information from HI interferometric surveys to estimate the HI scale 
height of a face--on galaxy.

\section{The SCF Method}

The Spectral Correlation Function (SCF) measures the correlation between 
spectra at different map positions as a function of their spatial separation, 
and is sensitive to the properties of the gas mass distribution and the gas 
velocity field (Rosolowsky et al. 1999; Padoan, Rosolowsky \& Goodman 2001).

Let $T(\vecr,v)$ be the antenna temperature as a function of velocity 
channel $v$ at map position $\vecr$. The SCF for spectra with spatial 
separation $\dr$ is: 

\begin{equation}
S_0(\dr)=\left\langle \frac{S_0(\vecr,\dr)}{S_{0,{\rm N}}(\vecr)} \right\rangle _{\vecr},
\label{1}
\end{equation}
where the average is done over all map positions $\vecr$. $S_0(\vecr,\dr)$ 
is the SCF uncorrected for the effects of noise,
\begin{equation}
S_0(\vecr,\dr)=\left\langle 1- \sqrt{\frac{\Sigma_v[T(\vecr,v)-T(\vecr+\vecdr,v)]^2}
{\Sigma_vT(\vecr,v)^2+\Sigma_vT(\vecr+\vecdr,v)^2}} \right\rangle _{\vecdr},
\label{2}
\end{equation}
where the average is limited to separation vectors $\vecdr$ with $|\vecdr|=\dr$,
and $S_{0,{\rm N}}(\vecr)$ is the SCF due entirely to noise,
\begin{equation}
S_{0,{\rm N}}(\vecr)=1-\frac{1}{Q(\vecr)},
\label{3}
\end{equation}
and $Q(\vecr)$ is the ``spectrum quality'' (see discussion in Padoan, Rosolowsky \& 
Goodman 2001). $Q(\vecr)$ is computed as the ratio of the rms signal within a 
velocity window $W$ to the rms noise, $N$,
\begin{equation}
Q(\vecr)=\frac{1}{N}\sqrt{\frac{\sum_v T(\vecr,v)^2dv}{W}},
\label{4}
\end{equation}
where $dv$ is the width of the velocity channels.

The SCF has been computed for both molecular cloud data and synthetic data obtained
by solving the radiative transfer through the three--dimensional density and velocity 
fields of numerical simulations of super--sonic MHD turbulence. 
The result is typically a power law that extends up to a separation $\dr$ comparable
to the map size, reflecting the self--similarity of super--sonic turbulence (see Padoan, 
Rosolowsky \& Goodman 2001, and references therein).

\ifnum\inlinefig=1
\begin{figure}[!th]
\centerline{\epsfxsize=8.2cm \epsfbox{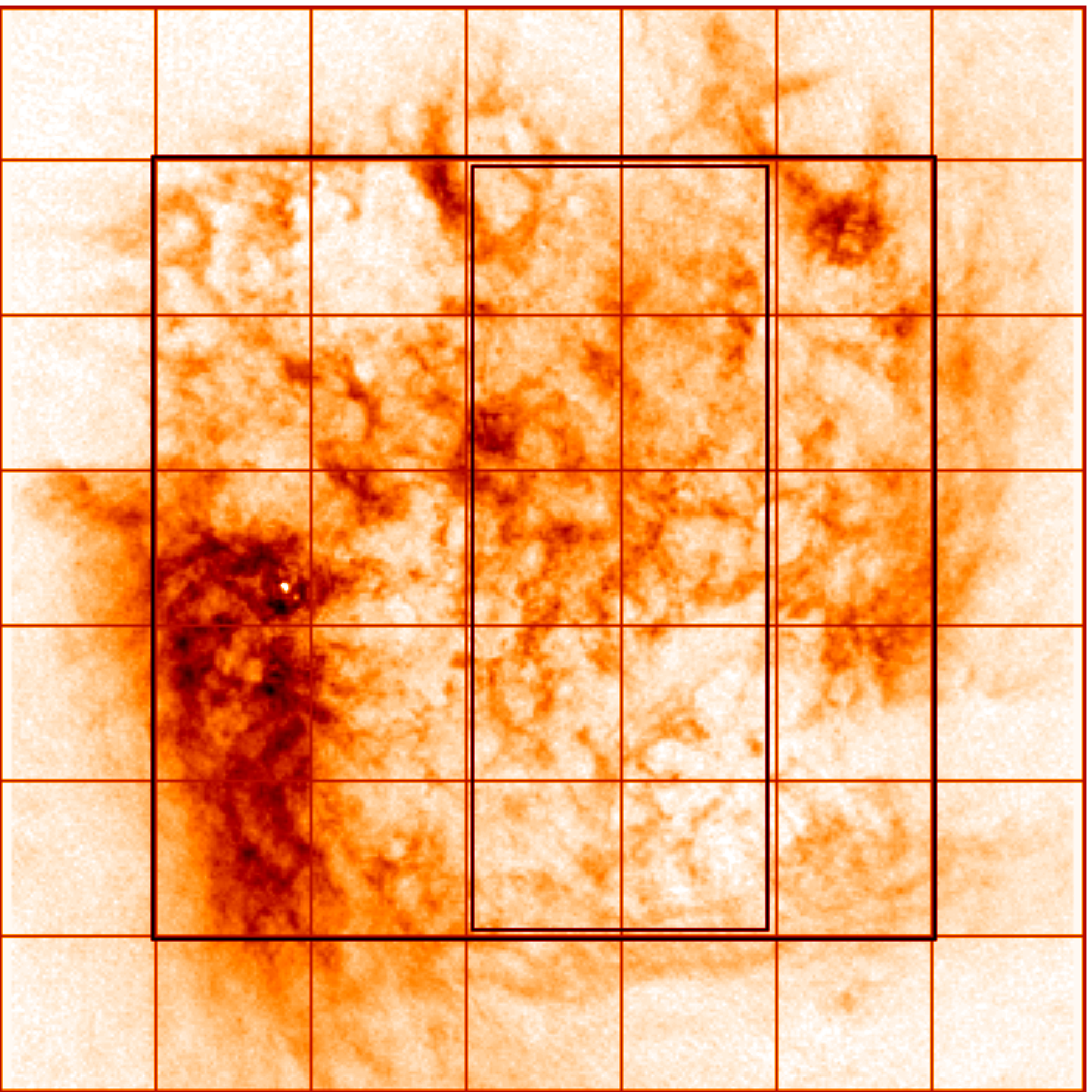}}
\caption[]{%
Velocity integrated HI intensity in the LMC from the ATCA 
interferometric survey by Kim et al. (1998a,b, 1999), combined 
with the single dish survey done with the Parkes Multibeam by 
Staveley--Smith et al. (in preparation). The perimeter of 
each of the 25 regions used to compute the SCF is over--plotted.}
\label{fig1}
\end{figure}
\fi

\ifnum\inlinefig=1
\begin{figure}[!th]
\centerline{\epsfxsize=8.7cm \epsfbox{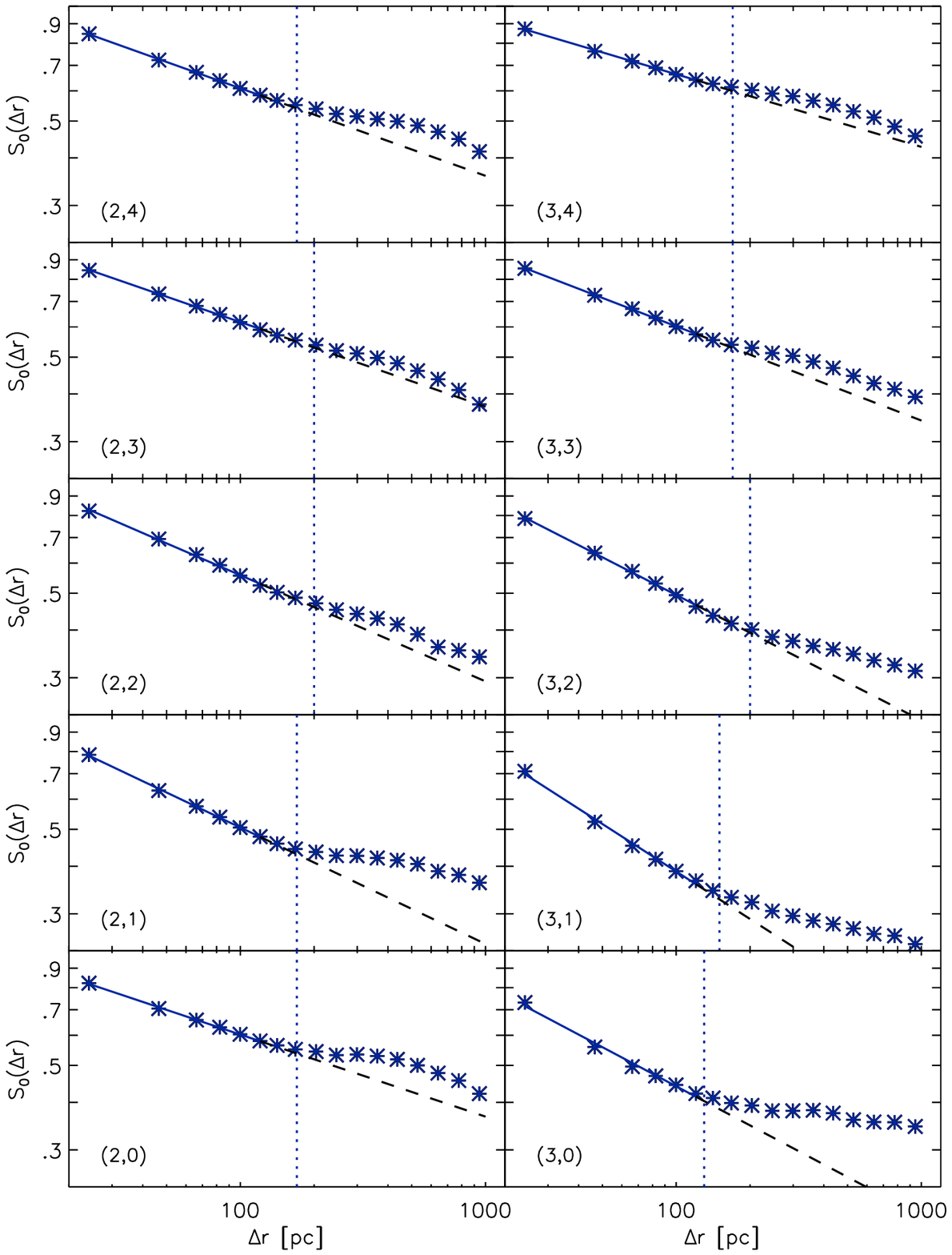}}
\caption[]{%
The SCF for the third and fourth column of regions from
Figure~1. The regions are given coordinates (x,y) starting from (0,0) at the 
bottom left corner of the $5\times 5$ square in Figure~1.}
\label{fig2}
\end{figure}
\fi

\ifnum\inlinefig=1
\begin{figure}[!th]
\centerline{\epsfxsize=8.2cm \epsfbox{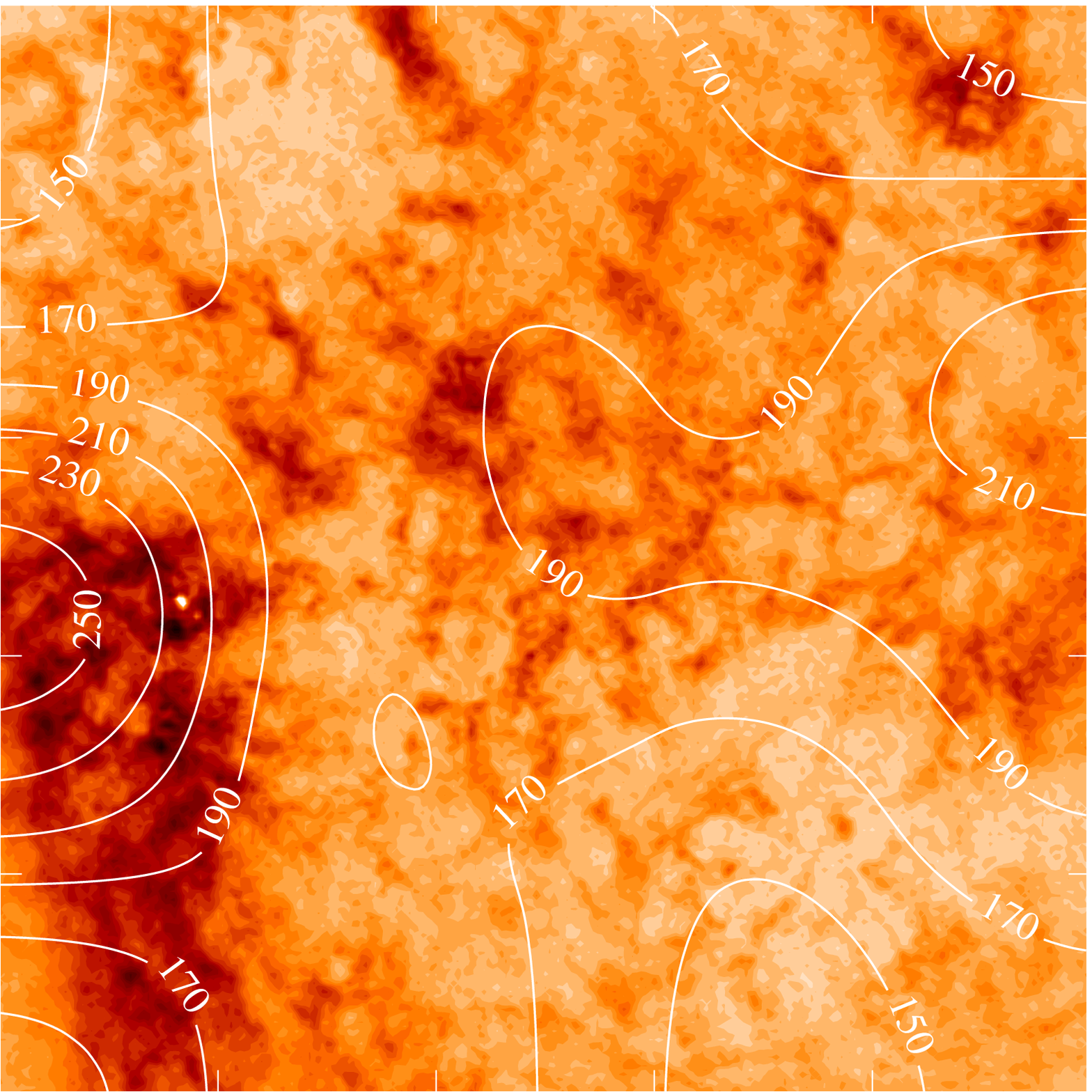}}
\caption[]{%
Contour map of the SCF scale height of the LMC disk. Values of contour 
lines are in pc. The intensity map shows the HI integrated intensity 
as in Figure~1.}
\label{fig3}
\end{figure}
\fi

\section{The SCF and the Scale--Height of the LMC HI Disk}

We have computed the SCF for several regions of the Australia Telescope 
Compact Array (ATCA) interferometric HI survey of the Large Magellanic Cloud 
(LMC) by Kim et al. (1998a,b, 1999), combined with the single dish survey 
done with the Parkes Multibeam by Staveley--Smith et al. (in preparation).
The original data has been rebinned to a resolution of 80'', that is a linear
resolution of 20~pc assuming a distance to the LMC of 50~kpc (Feast 1991). 
We therefore obtain a map made of $350\times350$ spectra ($7\times 7$~kpc), 
with 120 velocity channels. The velocity resolution is 1.65~km/s. Details 
about the ATCA survey can be found in Kim et al. (1998a,b, 1999).

\nocite{Feast91}

The finite size of a map with non--periodic boundaries generates edge effects
in correlation functions. In order to avoid this problem, we have
computed the SCF for subsets of $50\times 50$ spectra ($1\times 1$~kpc),
excluding the data from the ``edges'' of the map. The spectra 
in each subset have been compared with all spectra within a larger area around that 
subset, including $150\times 150$ spectra ($3\times 3$~kpc). In this way the SCF for 
all the spectra in each subset can be computed up to a separation equal to the size 
of the subset (1~kpc), without any effects related to the finite size of the map. 

Figure~1 shows the integrated HI intensity map of the LMC, divided into $7\times 7$
subsets with $50\times 50$ spectra each. The SCF has been computed only for the 
$5\times 5$ central ones (delimited by the darker square) for which complete sets
of $150\times 150$ reference spectra are available. For lack of space, we have 
plotted in Figure~2 only the SCF computed inside the area marked with the
rectangle in Figure~1.  

Figure~2 shows that the SCF is a power law on small scales over almost an order 
of magnitude range in spectral separation $\dr$. The power law slope of the SCF 
varies from -0.40 to -0.15 over the whole map. The power law is very clearly 
interrupted at $\approx 180$~pc, where the slope of the SCF generally decreases. 
In 22 of the 25 regions, the SCF breaks to a ``shallower'' slope on larger
scales. It sometimes becomes steeper again on the scale of a few hundred pc,
due probably to the effect of differential rotation, which grows with 
increasing spectral separation. In only 3 of the 25 cases, the SCF breaks from 
a power law to a ``steeper'' slope. This happens in the 3 regions around 
the location of 30 Doradus, and south of it (the 3 regions covering
the largest intensity feature on the bottom left of the map in Figure~1). 

We interpret the interruption of the power law shape of the SCF as the outer scale
of three dimensional super--sonic turbulence in the LMC disk. A scale--free 
cascade of three dimensional turbulence is unlikely to extend to scales much
larger than the disk scale height, as virtually two dimensional flows should 
dominate the dynamics on those larger scales. 
The power law shape of the SCF for the HI data reflects the presence
of self--similar turbulence, probably extending to much smaller scale than probed by 
the present survey (20~pc in our rebinned data). Below 20~pc the turbulent cascade
is well--probed by the SCF of molecular emission line maps (see Padoan, Rosolowsky 
\& Goodman 2001). 

Assuming that the the SCF is indeed sensitive to the HI scale height, $h$, we 
find that $h$ varies in the range $130<h<280$~pc over the surveyed area. 
The largest value is found in the region around 30 Doradus.
Statistical error bars are smaller than the size of the symbols used in the 
plots in Figure~2, and variations of $h$ are therefore significant. 
Figure~3 shows a contour map of $h$ over the $5\times 5$ regions where the 
SCF has been computed. The average value is $\langle h\rangle\approx 176$~pc. 
Kim et al. (1999) estimated the HI scale height using the average 
vertical velocity dispersion and the average surface density of both the HI 
and the stellar components of the LMC disk, and adopting the disk model by 
Dopita \& Ryder (1994). They found $h\approx 180$~pc. This is almost
identical to the value of $\langle h\rangle$ estimated with the SCF method,
which is the main reason why we interpret the outer correlation scale 
of the SCF as the local value of the disk scale height. 
Nevertheless, we cannot rule out possible alternative 
interpretations of the SCF results, and the subject deserves further study.

\nocite{Dopita+Ryder94}
 
Elmegreen, Kim \& Staveley--Smith (2001) have recently computed the power
spectrum of HI intensity maps of the LMC obtained with data from the same survey
used in this work. They find a steeper power spectrum on small scales than on 
large ones. The transition occurs at approximately 80--100~pc, and is interpreted 
as the line of sight thickness of the HI layer, based the theoretical results by
Lazarian \& Pogosian (2000).  

In this work we confirm the possibility of using the statistical information 
from HI interferometric surveys to estimate the HI scale height of a face--on 
galaxy. However, there is a fundamental difference between the present Letter 
and the work by Elmegreen, Kim \& Staveley--Smith (2001), apart from the 
different method of data analysis adopted. Elmegreen, Kim \& Staveley--Smith 
(2001) implicitly assume that the three dimensional power spectrum of the 
turbulent density field is described by a unique power law over all scales, 
and the change in slope of the two dimensional power spectrum of integrated 
intensity is merely a projection effect. Here instead we argue that there 
must be a physical transition in the statistical properties of the flow on 
a scale close to the disk scale height, which the SCF is sensitive to.

\section{Conclusions}

We have proposed a new way to measure and map the scale height of the gas
in face--on disk galaxies, based on the application of the SCF to HI interferometric
surveys. We have shown that this method applies successfully to the LMC.
It can be applied to a number of nearby galaxies, depending on the spatial and
velocity resolution of HI or molecular transition interferometric surveys.
The BIMA survey of nearby galaxies, BIMA SONG (Regan et al. 2001), 
for example, has an angular resolution
of approximately 6'' (using the D and C array configurations), and can achieve
approximately 3'', with the B array configuration. At the distance of 4 Mpc, 
a 3'' resolution corresponds to a linear resolution of $\approx 60$~pc. 
There are six galaxies within the approximate distance range between 2 and 4.5 Mpc, 
in the BIMA SONG list: IC0342, NGC2403, NGC2976, NGC3031, NGC4736 and NGC4826. 
They could in principle be studied with this method. Problems with the CO surveys 
may arise due to the low surface filling factor of CO relative to HI, and, in the 
particular case of the BIMA SONG survey, due to the relatively low velocity 
resolution (4.06~km/s). However, a velocity resolution of approximately
1.0~km/s could be easily achieved with BIMA in these galaxies (Tamara T. Helfer,
private communication). 

\nocite{Regan+2001}

For 21-cm HI observations, a 50-km baseline in a radio interferometer gives
a resolution of 16 pc at 4 Mpc. At 3-mm, CO gives the same kind of resolution 
on a 0.7-km baseline. Therefore, the synthesized beams of the immediate descendants 
of existing cm and mm interferometers (e.g. the EVLA and CARMA) should be able to map 
many galaxies with more--than--adequate resolution for the SCF to be used to measure, 
and map out, galactic scale height. 

The application of this method to a number of galaxies could provide new important
measurements of the gas scale height and its variations inside individual disks and 
between different galaxies. Once the scale height is measured, it can also be used, 
together with the observed vertical velocity dispersion, to put new constraints on 
the distribution of dynamical mass in galactic disks.

\acknowledgements

We are grateful to Jos\'{e} Franco for helpful comments.
This work was supported by NSF grant AST-9721455.



\ifnum\inlinefig=0
\clearpage

\onecolumn

{\bf Figure captions:} \\

{\bf Figure \ref{fig1}:} Velocity integrated HI intensity in the LMC from the ATCA 
interferometric survey by Kim et al. (1998a,b, 1999), combined 
with the single dish survey done with the Parkes Multibeam by 
Staveley--Smith et al. (in preparation). The perimeter of each of the 
25 regions used to compute the SCF is over over--plotted. \\

{\bf Figure \ref{fig2}:} The SCF for the third and fourth column of regions from
Figure~1. The regions are given coordinates (x,y) starting from (0,0) at the bottom
left corner of the $5\times 5$ square in Figure~1. \\

{\bf Figure \ref{fig3}:} Contour map of the SCF scale height of the LMC disk. Values of 
contour lines are in pc. The intensity map shows the HI integrated intensity as in Figure~1.\\

\clearpage
\begin{figure}
\centerline{\epsfxsize=15cm \epsfbox{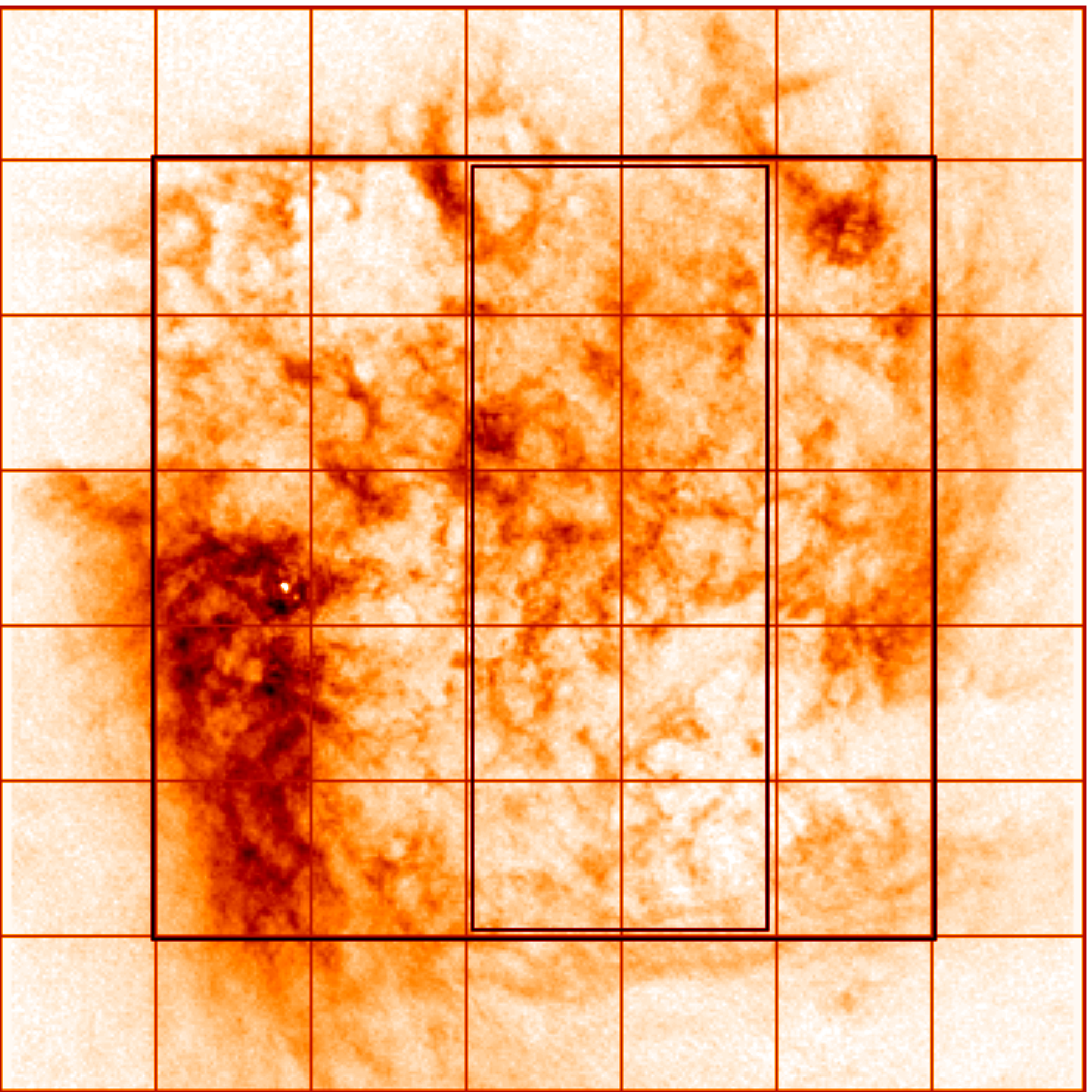}}
\caption[]{}
\label{fig1}
\end{figure}

\clearpage
\begin{figure}
\centerline{\epsfxsize=15cm \epsfbox{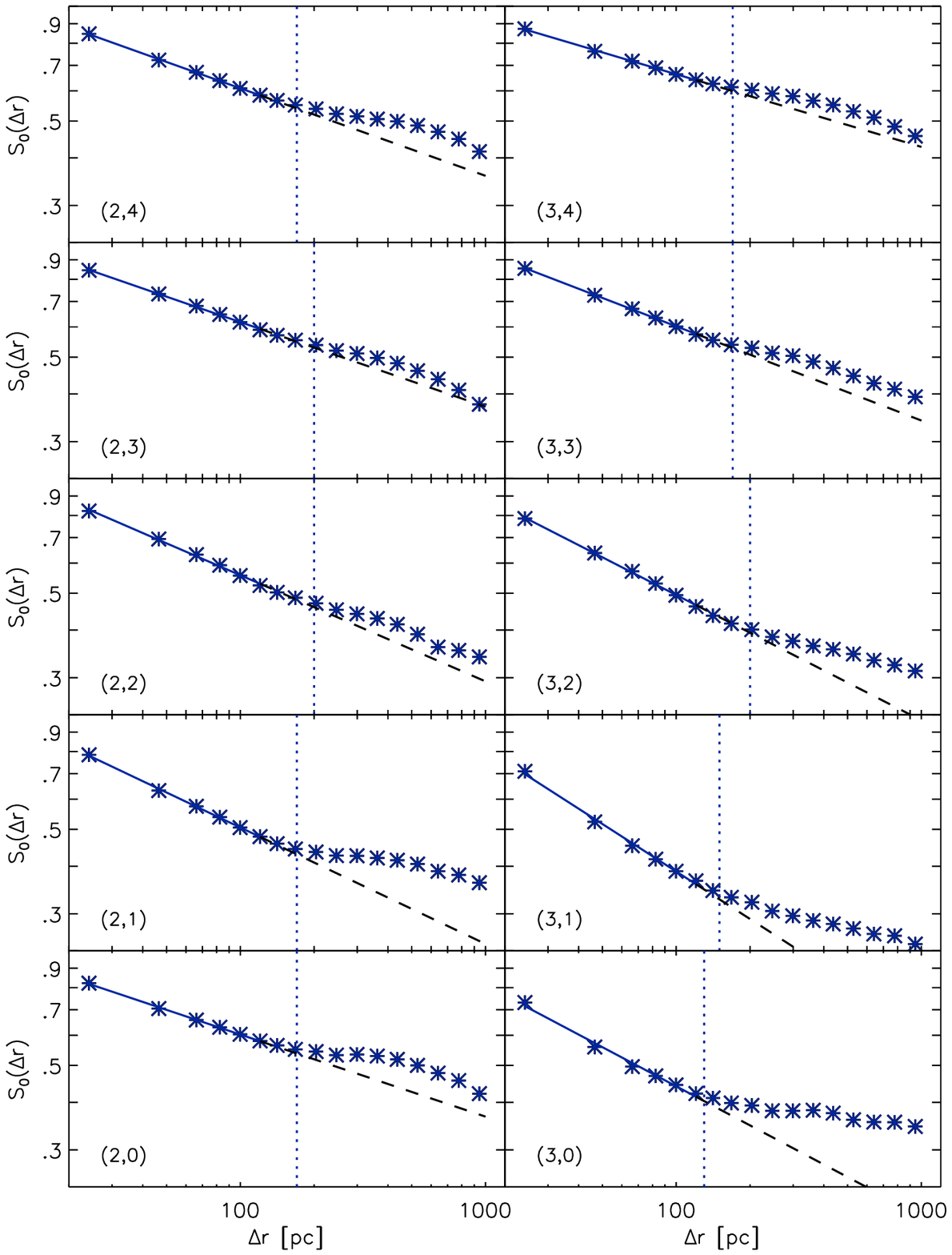}}
\caption[]{}
\label{fig2}
\end{figure}

\clearpage
\begin{figure}
\centerline{\epsfxsize=15cm \epsfbox{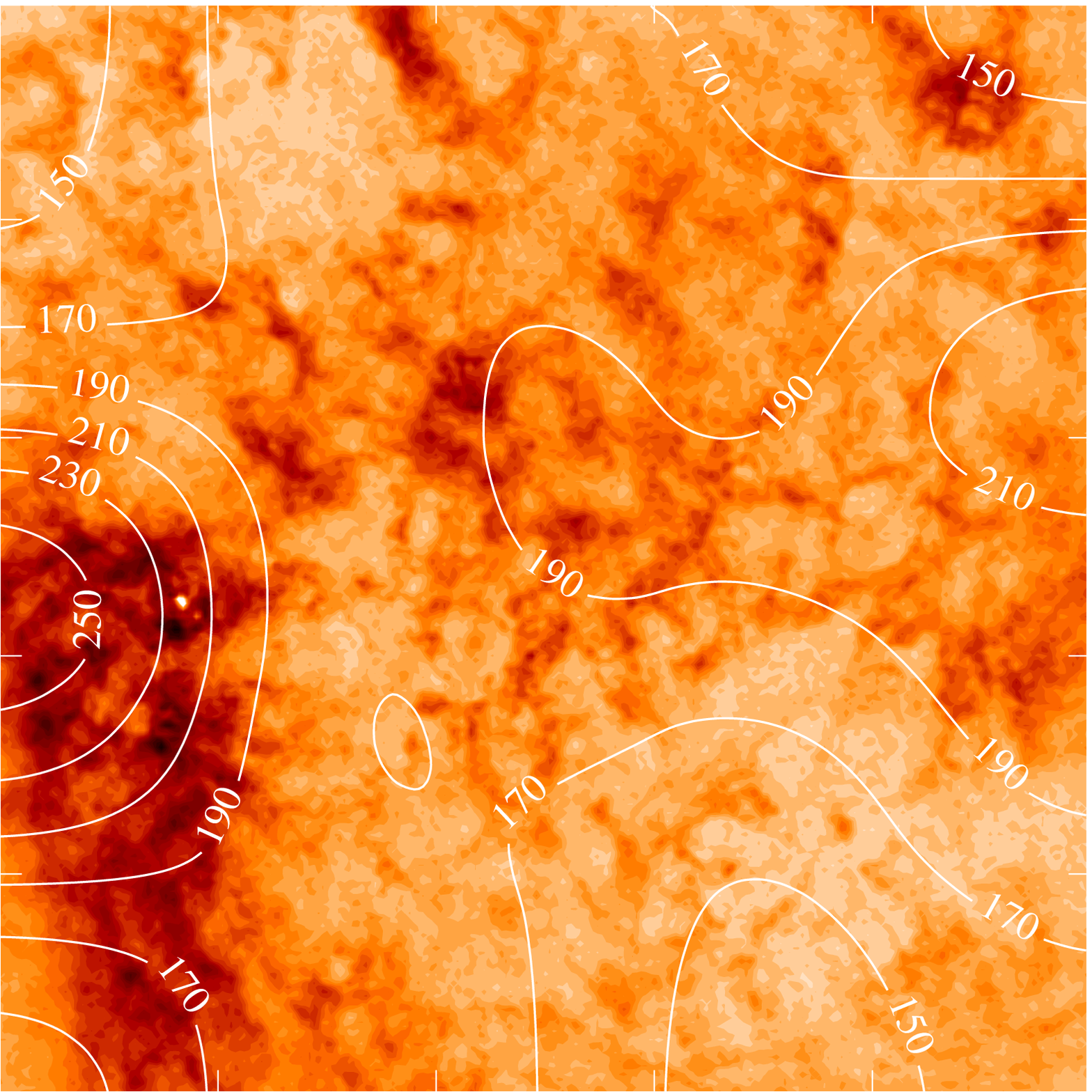}}
\caption[]{}
\label{fig3}
\end{figure}

\fi

\end{document}